\documentclass{aa}

\usepackage{graphicx}
\usepackage{txfonts}
\usepackage[pdfpagelabels=false]{hyperref}
\hypersetup{colorlinks=true,linkcolor=blue,citecolor=blue,filecolor=blue,urlcolor=blue,}
\usepackage[dvipsnames]{xcolor}
\newcommand{\new}[1]{\textcolor{OliveGreen}{#1}}

\begin{document}

   \title{The binary content of multiple populations in NGC~3201}


   \author{S. Kamann\inst{1}
           \and
           B. Giesers\inst{2}
           \and           
           N. Bastian\inst{1}
           \and
           J. Brinchmann\inst{3,4}
           \and
           S. Dreizler\inst{2}
           \and
           F. G\"ottgens\inst{2}
           \and
           T.-O. Husser\inst{2}
           \and
           M. Latour\inst{2}
           \and
           P.~M. Weilbacher\inst{5}
           \and
           L. Wisotzki\inst{5}
          }

   \institute{
    Astrophysics Research Institute, Liverpool John   Moores University, 146 Brownlow Hill, Liverpool L3 5RF,  UK\\          \email{s.kamann@ljmu.ac.uk}
    \and
    Institute for Astrophysics, Georg-August-Universit\"at G\"ottingen, Friedrich-Hund-Platz 1, 37077 G\"ottingen, Germany
    \and
    Instituto de Astrof{\'\i}sica e Ci{\^e}ncias do Espa\c co, Universidade do Porto, CAUP, Rua das Estrelas, PT4150-762 Porto, Portugal
    \and
    Leiden Observatory, Leiden University, P.O. Box 9513, 2300 RA, Leiden, The Netherlands
    \and
    Leibniz-Institute for Astrophysics, An der Sternwarte 16, 14482 Potsdam, Germany
             }

   \date{Received Oct. 4, 2019, }

 
  \abstract{We investigate the binary content of the two stellar populations that coexist in the globular cluster NGC~3201. Previous studies of binary stars in globular clusters have reported higher binary fractions in their first populations (P1, having field-like abundances) compared to their second populations (P2, having anomalous abundances). This is interpreted as evidence for the latter forming more centrally concentrated. In contrast to previous studies, our analysis focuses on the cluster centre, where comparable binary fractions between the populations are predicted because of the short relaxation times. However, we find that even in the centre of NGC~3201, the observed binary fraction of P1 is higher, $(23.1\pm6.2)\%$ compared to $(8.2\pm3.5)\%$ in P2. Our results are difficult to reconcile with a scenario where the populations only differ in their initial concentrations, but instead suggests that the populations also formed with different fractions of binary stars.}


   \maketitle
%

\section{Introduction}

One of the lesser studied aspects of the multiple populations \citep[a.k.a. abundance anomalies, see][for a review]{2018ARA&A..56...83B} phenomena in massive stellar clusters is the role of stellar binarity.  This is due to the overall relatively low binary fractions in globular clusters \citep[GCs, e.g.,][]{2015ApJ...807...32J} and the fact that it is difficult to separate out the binaries from (apparent) single stars in colour-magnitude diagrams for each of the populations (i.e., the ``normal'' and ``anomalous'' stars; P1 and P2) as the sequences overlap.  Instead, one must carry out intensive spectroscopic time-series analyses of a representative sample of stars from each population to search for radial velocity variations.

The most comprehensive survey using this technique, to date, was that of \citet{2015A&A...584A..52L} who monitored 968 red giant branch (RGB) stars in ten Milky Way ancient GCs.  From this large sample they found 21 binary stars and when separating their sample into P1 and P2 stars, found binary fractions of $4.9$\% and $1.2$\% for each population, respectively. In addition, \citet{2018ApJ...864...33D} recently reported a higher binary fraction in P1 of the globular cluster NGC~6362, $14\%$ compared to $<1\%$ in P2.

Such differences can be explained in terms of the formation environment of the stars, with the P2 stars (lower binary fraction) forming and initially evolving in a much denser environment, which would destroy many of the primordial binaries \citep[e.g.][]{2016MNRAS.457.4507H}. An initially more concentrated P2 is a common feature of essentially all scenarios put forward to explain multiple populations and appears to be in agreement with the observed density profiles and kinematics of the populations in most Galactic globular clusters today \citep[e.g.][]{2011A&A...525A.114L,2013ApJ...771L..15R,2015ApJ...810L..13B,2019ApJ...884L..24D}.

Due to the fibre based observations, the targets for the study of \citet{2015A&A...584A..52L} were preferentially located in the outer regions of the clusters. Most globular clusters show a trend of increasing binary fractions towards the cluster centres \citep[e.g][]{2012A&A...540A..16M}, which is thought to be due to mass segregation. On the other hand, dynamical processes lowering the binary fractions, such as binary disruption or ejection from the cluster, occur more frequently near the cluster centres. Therefore, the binary statistics near the cluster centres may not follow those in the cluster outskirts. Using N-body simulations, \citet{2015MNRAS.449..629H,2016MNRAS.457.4507H} found that the binary fractions of P2 are expected to be comparable or even larger than those of P1 inside the clusters' half-light radii if P2 formed centrally concentrated.

In the present work we explicitly test these predictions, using the time series VLT/MUSE observations of NGC~3201 stretching over $>4\,{\rm years}$ presented in \citet{2019arXiv190904050G}, which focus on the region inside the core radius \citep[$r_{\rm c}=1.3\arcmin\equiv1.85\,{\rm pc}$,][]{1996AJ....112.1487H} of the cluster. RGB stars from the different populations are found using a UV-optical ``chromosome map'' \citep{2017MNRAS.464.3636M} which is highly efficient in separating the populations, largely based on their N abundance differences \citep[][]{2018A&A...616A.168L}.

\section{Data}

\begin{figure*}
    \begin{minipage}{.5\textwidth}
    \includegraphics[width=\columnwidth]{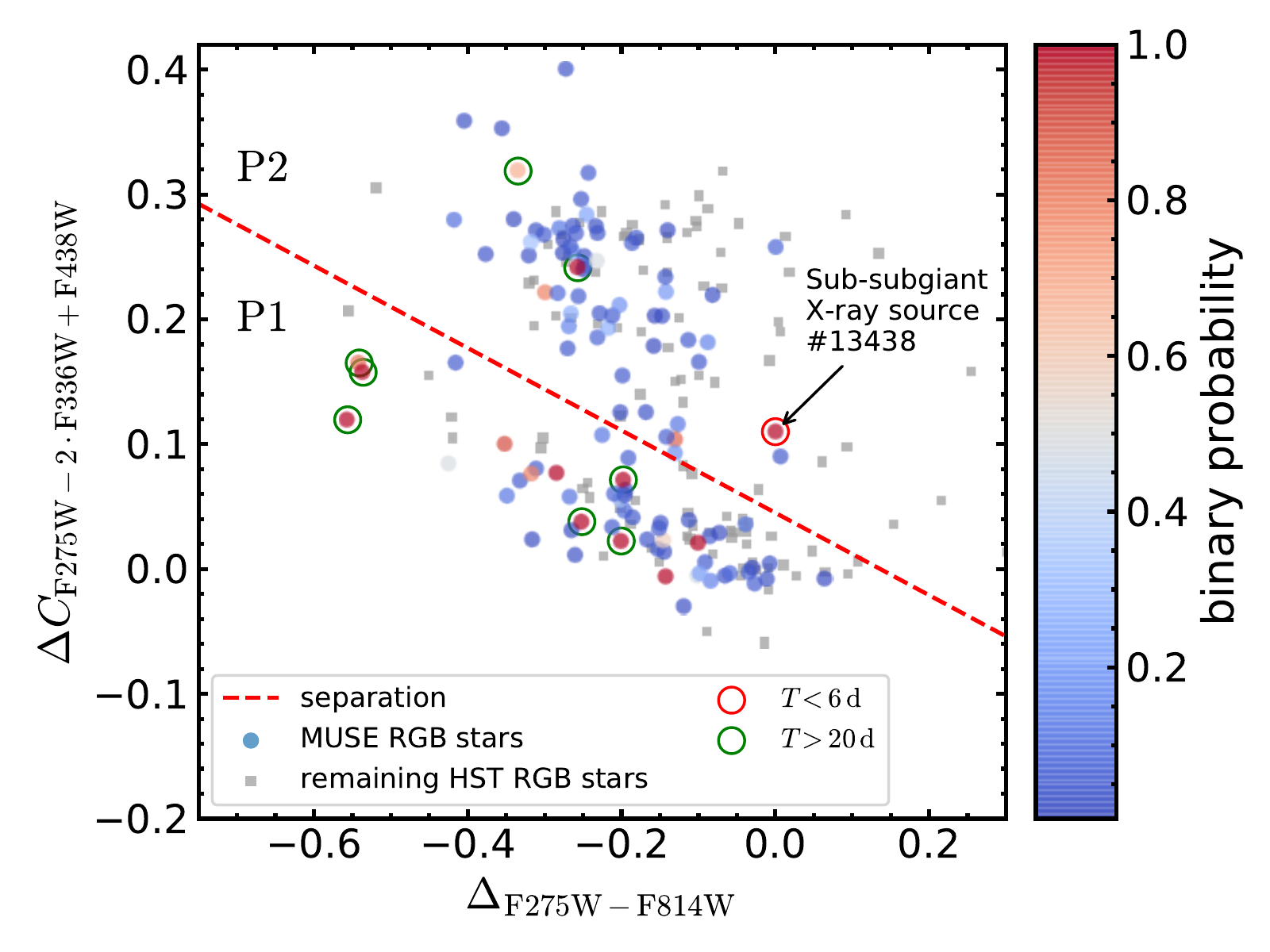}
    \end{minipage}
    \begin{minipage}{.5\textwidth}
    \includegraphics[width=\columnwidth]{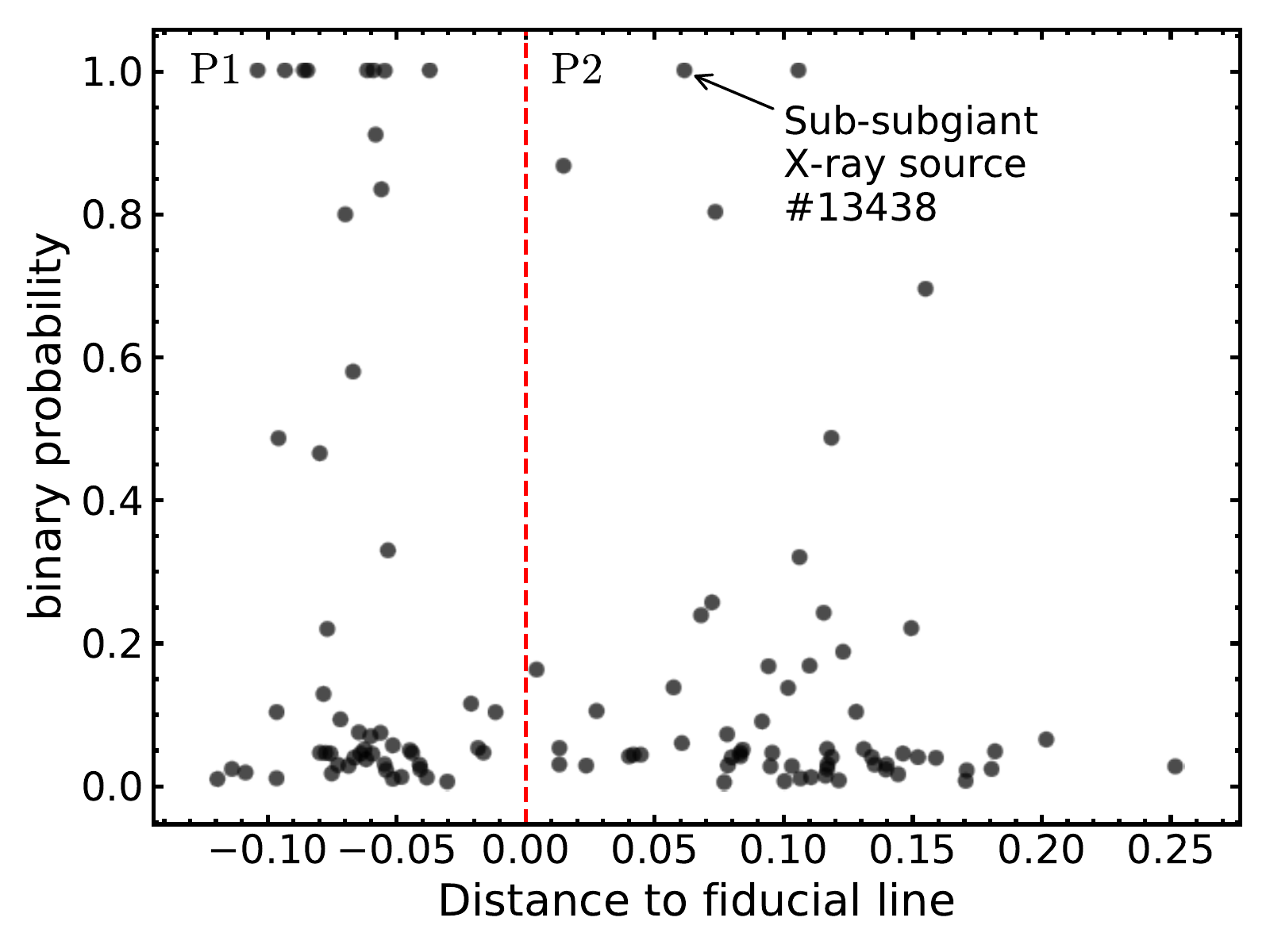}
    \end{minipage}
    \caption{\textit{Left:} The distribution of red giant stars in NGC~3201 in pseudo-color space -- so-called ``chromosome map''. The full sample of stars obtained from the \textit{HST} photometry of \citet{2015AJ....149...91P} is shown as small grey dots. Stars available in the MUSE sample are highlighted and colour-coded according to their probability of being in a binary system. For the sub-sample of MUSE sources with known Keplerian parameters, coloured rings indicate the orbital period $T$. The dashed red line illustrates our separation into P1 and P2 stars. \textit{Right:} The binary probability of the stars in the MUSE sample is shown as a function of the distance of a star perpendicular to the line separating P1 and P2 (i.e. the dashed red line in the left panel).}
    \label{fig:chromosome_map}
\end{figure*}

NGC~3201 has been observed as part of the MUSE survey of Galactic globular clusters \citep[see][]{2018MNRAS.473.5591K}, a large GTO programme targeting the central regions of massive star clusters. To facilitate the detection and characterization of binary stars, repeated observations of five pointings, covering approximately the central $2\arcmin\times2\arcmin$ of the cluster, have been performed from November 2014 to May 2019. The data analysis, including the detection and characterization of binaries, are described in \citet{2019arXiv190904050G}. For each of the $3\,553$ stars studied, the authors provided a probability that the star shows radial velocity variations. The radial velocities of the stars with at least 5 observations and with a probability higher than $50\%$ of being variable were further analysed with \textsc{The Joker} \citep{2018AJ....156...18P}, resulting in a subset of $95$ stars with unique Keplerian orbit solutions.

To split up the two stellar populations that have previously been identified in NGC~3201, we used the \textit{Hubble Space Telescope} photometry from the survey of \citet[][see \citealt{2018MNRAS.481.3382N}]{2015AJ....149...91P}. As outlined in \citet{2019A&A...631A..14L}, this was done by creating a ``chromosome map'' from the red giant stars, which is shown in Fig.~\ref{fig:chromosome_map}. The separation of the two populations has been performed following \citet{2017MNRAS.464.3636M} and is indicated by the red dashed line included in Fig.~\ref{fig:chromosome_map} (with P1 being below the fiducial line).

Finally, we identified the subset of stars from \citet{2019arXiv190904050G} for which the population could be determined. This resulted in a final sample of $113$ stars, $52$ in P1 and $61$ in P2, that is presented in Table~\ref{tab:app:muse_rgb_sample}. For the $17$ out of $113$ stars which had a probability $P>0.5$ of being in a binary and sufficient ($\geq5$) observations, we tried to determine the Keplerian orbit. This resulted in a subset of $9$ stars, for which an orbit solution is available. The orbital parameters of said stars are included in Table~\ref{tab:app:muse_rgb_sample}. The remaining $8$ stars with $P>0.5$ have insufficient kinematical data to infer their Keplerian orbits.

\section{Results}

\subsection{Binaries across the chromosome map}

The distribution of binary stars across the chromosome map of NGC~3201 is shown in the left panel of Fig.~\ref{fig:chromosome_map}. We colour-coded each star available in the sample of \citet{2019arXiv190904050G} by its probability to be in a binary system. Stars for which orbital solutions have been found are further highlighted according to their orbital period $T$. To better visualize possible differences between P1 and P2, we show in the right panel of Fig.~\ref{fig:chromosome_map} the binary probability as a function of the distance perpendicular to the fiducial line separating P1 and P2.

To infer the binary fractions in both populations, we follow \citet{2019arXiv190904050G} and obtain the fraction of stars with a binary probability of $P>0.5$ within each population. This leads to binary fractions of $(23.1\pm6.2)\%$ in P1 and ($8.2\pm3.5)\%$ in P2. The uncertainties tailored to both values take into account the limited sample sizes as well as the uncertainties stemming from the threshold in $P$ \citep[see][for details]{2019arXiv190904050G}. When calculating the binary fraction in P2, we included the sub-subgiant star highlighted in Fig.~\ref{fig:chromosome_map}, which is in a much tighter orbit than the remaining binary systems (indicated by the coloured rings in Fig.~\ref{fig:chromosome_map}). As discussed in \citet{2019arXiv190904050G}, this star has an X-ray counterpart and shows \ion{H}{$\alpha$} emission. Hence it is plausible that this star is part of an accreting binary system, which would also impact its photometric properties and its location in the chromosome map. Excluding it from our calculation reduces the binary fraction of P2 to $(6.7\pm3.3)\%$. Averaged over both populations, we find a binary fraction of $(15.0\pm3.4)\%$, in good agreement with the discovery fraction of $(17.1\pm1.9)\%$ determined by \citet{2019arXiv190904050G}.

\subsection{The origin of the observed binaries}

To study the origin of the observed binaries, we make use of the subsample with known orbital parameters. The fate of a binary in a globular cluster is linked to its hardness $h$, i.e. the ratio of its internal energy  $\Tilde{E}$ to the average kinetic energy of the surrounding stars,
\begin{equation}
    h = \lvert \Tilde{E} \rvert / m \sigma^2\,,
\end{equation}
where $m$ is the typical stellar mass of a cluster member and $\sigma$ the velocity dispersion of the cluster. For a bound Keplerian orbit, the internal energy is given as 
\begin{equation}
    \Tilde{E} = -\frac{G m_p m_c}{2a}\,,
\end{equation}
with $m_p$ and $m_c$ being the masses of the constituents and $a$ being the semi-major axis of the binary. In Table~\ref{tab:app:muse_rgb_sample}, we provide the hardness for each binary with an orbit available. The values were calculated assuming an inclination of $i=90^{\circ}$, i.e. the minimum possible companion mass $m_c$. The mass of the primary (RGB) component, $m_p$, was determined via comparison to an isochrone as described in \citet{2019arXiv190904050G}. As cluster properties, we used $m=0.8\,{\rm M_{\odot}}$ and $\sigma=4.3\,{\rm km\,s^{-1}}$ \citep{2018MNRAS.478.1520B}. All systems are hard binaries with $h>1$, indicating that they can survive in NGC~3201 for a Hubble time. Their longevity can be confirmed by determining the expected lifetimes of the binary stars, $\tau=1/B(\Tilde{E})$, where $B(\Tilde{E})$ is the probability of a binary being ionized (i.e. destroyed) in a gravitational encounter with a third cluster member, given as \cite[eq.~7.174 in][]{2008gady.book.....B}
\begin{equation}
    B(\Tilde{E}) = \frac{8\sqrt{\pi}G^2m^3\rho\sigma}{3^{3/2}\lvert \Tilde{E} \rvert}\left(1 + \frac{1}{5 h} \right)^{-1}\left[1+ e^h\right]^{-1}\,.
    \label{eq:ion_prob}
\end{equation}
Evaluating eq.~\ref{eq:ion_prob} for a core density of $\rho=10^{2.72}\,{\rm M_{\odot}/pc^3}$ \citep{2018MNRAS.478.1520B} yields lifetimes for all binary stars that exceed the age of NGC~3201 by several orders of magnitude.

Note that the companion masses $m_{2}$ and the semi-major axes $a$ used in the above calculations were derived under the assumption that the binaries are observed edge-on (i.e. at an inclination of $i=90\,{\rm deg}$). While both quantities increase with decreasing inclinations, $m_{2}$ is more sensitive on $i$ than $a$ is, so that our hardness values can be considered as lower limits.

Finally, we stress that the probability to form hard binaries dynamically in a relatively low-density cluster such as NGC~3201 is very small. Using eq.~7.176 from \citet{2008gady.book.....B}, the formation rate of hard binaries per unit volume is given as
\begin{equation}
    C_{\rm hb} = 0.74\frac{G^5\rho^3 m^2}{\sigma^9}\,.
    \label{eq:formation_rate}
\end{equation}
Integrating eq.~\ref{eq:formation_rate} over the core of NGC~3201 \citep[assuming $r_{\rm c}=1.74\,{\rm pc}$,][]{2018MNRAS.478.1520B} yields a total formation rate of $7\times10^{-6}/{\rm Gyr}$. Hence it is very likely that all binary stars that we observe in NGC~3201 are primordial.

\subsection{The impact of the companion}
\label{sec:companion}

Our determination of the binary fraction in each population is based on the assumption that the positions of the stars in the chromosome map are not altered by the presence of their companions. To verify this assumption, we used the binaries with known orbits and inferred the magnitude changes caused by their companions in the four HST filters underlying the chromosome map, F275W, F336W, F438W, and F814W. To this aim, we fetched an isochrone tailored to the properties of NGC~3201 \citep[${\rm [Fe/H]}=-1.59$, $E_{\rm B-V}=0.24$,][]{1996AJ....112.1487H} from the MIST database \citep{2016ApJ...823..102C}. We made the assumption that the companions are main sequence stars and predicted their magnitudes ${\rm mag_c}$ by selecting the isochrone points along the main sequence closest to their measured masses, $m_{\rm c}\sin i$ (cf. Table~\ref{tab:app:muse_rgb_sample}). As the measured companion masses need to be corrected for the (unknown) orbit inclinations relative to the line of sight, we assumed different inclination angles $i$ and found the isochrone counterpart for each value of $m_{\rm c}$. At each inclination, we calculated the corrected magnitude ${\rm mag_p}$ of the RGB star in the four relevant filters, according to
\begin{equation}
    {\rm mag_p} = {\rm mag_{tot}}-2.5\log_{10}\left(1 - 10^{-0.4({\rm mag_c} - {\rm mag_{tot}})}\right)\,,
\end{equation}
where ${\rm mag_{tot}}$ is the measured magnitude of the system in the considered filter. Then we predicted the actual location of each RGB star in the chromosome map. We stopped when subtracting the contribution of a fiducial companion resulted in a predicted position that was off by more than $0.1\,{\rm mag}$ from the red edge of the red giant branch in either (${\rm F275W}-{\rm F814W}$) colour or (${\rm F275W} - 2\cdot{\rm F336W} + {\rm F438W}$) pseudo-colour.

\begin{figure}
	\includegraphics[width=\columnwidth]{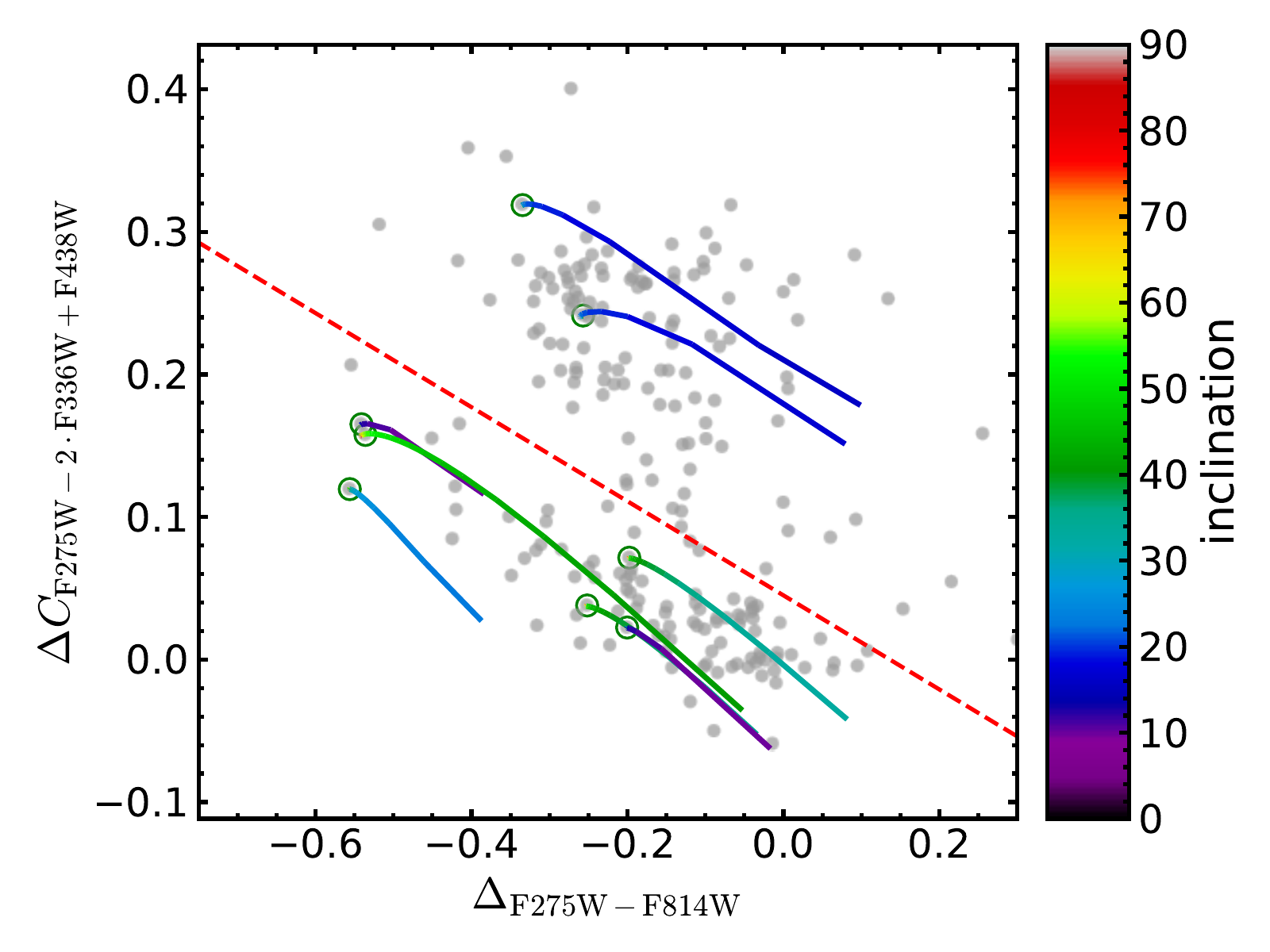}
    \caption{Impact of binary companions on the positions of RGB stars in the chromosome map of NCG~3201. For each star from our sample with known orbit, we show the true position of the RGB star after subtracting the contribution of a main sequence companion, as a function of orbit inclination. The grey points and the dashed red line are the same as in the left panel of Fig.~\ref{fig:chromosome_map}.}
    \label{fig:chromosome_shifts}
\end{figure}

We summarize the outcome of this test in Fig.~\ref{fig:chromosome_shifts}. It shows the predicted position of the RGB star in the chromosome map as a function of the inclination for each binary in our sample with a known orbit. Fig.~\ref{fig:chromosome_shifts} shows that P2 stars in a binary with a main sequence star are very unlikely to appear as P1 stars (and vice versa), as the companion tends to shift the binary in a direction parallel to the fiducial line separating the populations.

We further find that the companion needs to be massive enough to appear close to the main sequence turn-off in order to have a significant effect. Fig.~\ref{fig:chromosome_shifts} shows that for all of the sources in our sample, their orbits would need to be observed at low inclinations, $i \lesssim 40^{\circ}$, in that case, because our minimum masses are significantly below the expected turn-off mass of NGC~3201 ($m_{\rm TO}\sim0.8\,{\rm M_\odot}$, cf. Table~\ref{tab:app:muse_rgb_sample}). Under the assumption of randomly oriented orbits, we can estimate the probabilities to observe the systems at or below the inclinations where the companions have a measurable effect on the observed positions in the chromosome map. We find probabilities between $<1\%$ and about $35\%$ for the individual systems. Considering the sum of probabilities for the eight stars, we expect about one star among them which has been measurably shifted by its companion. Note that these probabilities do not account for the selection bias of radial velocity studies, which are more sensitive to edge-on orbits, and hence can be considered as upper limits \citep[see discussion in][]{2006ima..book.....C}.

We also considered the possibility of white-dwarf companions, as they may have a stronger impact on the F275W flux. However, in the photometry of \citet{2018MNRAS.481.3382N}, we find only $10-15$ white dwarf candidates with a F275W magnitude within $2~{\rm mag}$ of the main sequence turn-off.

\section{Discussion}

At first glance, our finding of a lower binary fraction in P2 than in P1 agrees with previous studies on the binary content of multiple populations \citep{2015A&A...584A..52L,2018ApJ...864...33D}. This trend was attributed to the P2 stars forming centrally concentrated, resulting in a higher rate of binary ionization and ejection. However, in contrast to earlier studies, our observations focus on the dense cluster core. In the simulations of \citet{2015MNRAS.449..629H,2016MNRAS.457.4507H}, the overabundance in P1 binaries typically only develops outside the half-light radii of the simulated clusters, whereas the trend disappears or even reverses inside of it.\footnote{Note that in contrast to \citet{2015MNRAS.449..629H,2016MNRAS.457.4507H}, we can only infer the binary fractions as a function of projected radius.} The observation by \citet{2018ApJ...864...33D} that the discrepancy in the observed velocity dispersions between P1 and P2 stars in NGC~6362 -- which is attributed to the overabundance of P1 binaries -- disappears towards the centre can also be interpreted as a hint towards comparable central binary fractions.

Compared to the clusters simulated by \citet{2015MNRAS.449..629H,2016MNRAS.457.4507H}, where all stars were input with the same masses, NGC~3201 appears much more complex. One complication is the likely presence of a large population of stellar-mass black holes \citep{2018MNRAS.475L..15G,2019arXiv190904050G,2018MNRAS.478.1844A}, which is expected to have a strong impact on the evolution of NGC~3201. Owing to their masses, the black holes can efficiently suppress the segregation of the binaries to the cluster centre. As the evolution of a binary population is governed by the interplay between mass segregation and their interactions with other stars, this marks an important difference compared to the existing simulations. Dedicated simulations using a realistic range of stellar masses will be an important step forward towards understanding the evolution of binary stars in multiple populations.

A possible explanation for our results is that P2 had different binary properties than P1 upon formation, e.g. a lower primordial binary fraction or a different distribution of semi-major axes. Most formation scenarios predict P2 stars to form while at least part of the P1 population is already in place. It seems likely that such vastly different formation environments had an impact on the properties of the primordial binaries in P2. Future hydrodynamical simulations of cluster formation may be able to investigate this further.

We note that some of the detected P1 binaries appear in a region of the chromosome map that is termed the \emph{extended P1} \citep[e.g.][]{2018A&A...616A.168L}, i.e. to the top-left of the bulk of P1 stars. As shown by, e.g., \citet{2019MNRAS.485.4128C}, extended P1 stars show no differences in their abundances of C, N, O, Na, Mg or Al compared to normal P1 stars. Very recently, \citet{2019arXiv191002892M} argued that binaries could be responsible for creating extended P1 stars in NGC~3201. As our analysis of Sect.~\ref{sec:companion} shows, normal P1 stars in binary systems with main sequence stars close to the turn-off can be shifted into the extended branch. However, it appears unlikely that this scenario is responsible for all of the stars observed along the extended P1. The binary systems with unique orbital solutions would need to be observed at rather unlikely inclination angles for the companions to produce noticeable shifts. In addition, a number of extended P1 stars do not show any signs of variability in our sample. Nevertheless, extending our analysis to other clusters with a pronounced extended P1, such as NGC~2808, appears as a very promising step in studying the impact of binary stars on the distribution of stars across chromosome maps.

Finally, we note that in comparison to \citet{2019arXiv191002892M}, our extended P1 extends to smaller values of $\Delta_{\rm F275W - F814W}$. Upon removal of the stars with $\Delta_{\rm F275W - F814W} \lesssim -0.4$, which were not considered in the work by \citet{2019arXiv191002892M}, our P1 binary fraction reduces to $19.1\pm5.7\%$. Hence our main conclusion of a higher binary fraction in P1 than in P2 does not depend on the exact definition of which stars belong to P1.

\begin{table*}
\longtab{
\begin{longtable}{ccccccccccc}
\caption{\label{tab:app:muse_rgb_sample} Photometric and orbital properties of the binaries with unique Kepler solutions in the MUSE sample. For each star, we provide the ID in the photometric catalogue of \citet{2008AJ....135.2055A}, the location in the chromosome map, the population tag, the binary probability, the mass of the primary star, the minimum mass of the companion star, the semi-major axis, \new{eccentricity, and period of the orbit}, and its hardness.}\\
\hline \hline
ACS Id & $\Delta_{\rm 275,\,814}$ & $\Delta C_{\rm 275,\,336,\,438}$ & pop. & $P_{\rm bin}$ & $m_{\rm p}/{\rm M_\odot}$ & $m_{\rm c}\sin i/{\rm M_\odot}$ & $a/{\rm AU}$\tablefootmark{a} & $e$ & $T/{\rm d}$ & $h$ \\

\hline
\endfirsthead
\caption{continued.}\\
\hline\hline
ACS Id & $\Delta_{\rm 275,\,814}$ & $\Delta C_{\rm 275,\,336,\,438}$ & pop. & $P_{\rm bin}$ & $m_{\rm p}/{\rm M_\odot}$ & $m_{\rm c}\sin i/{\rm M_\odot}$ & $a/{\rm AU}$\tablefootmark{a} & $e$ & $T/{\rm d}$ & $h$ \\

\hline
\endhead
\hline
\endfoot
3092 & 0.064 & -0.008 & 1 & 0.006 &  &  &  &  &  &  \\
3248 & 0.000 & 0.257 & 2 & 0.065 &  &  &  &  &  &  \\
3795 & -0.142 & 0.222 & 2 & 0.188 &  &  &  &  &  &  \\
4121 & -0.424 & 0.084 & 1 & 0.486 &  &  &  &  &  &  \\
4125 & -0.320 & 0.251 & 2 & 0.027 &  &  &  &  &  &  \\
4566 & -0.225 & 0.107 & 1 & 0.103 &  &  &  &  &  &  \\
4698 & -0.243 & 0.317 & 2 & 0.049 &  &  &  &  &  &  \\
4853 & -0.119 & -0.030 & 1 & 0.019 &  &  &  &  &  &  \\
5281 & -0.348 & 0.059 & 1 & 0.103 &  &  &  &  &  &  \\
5461 & -0.201 & 0.125 & 2 & 0.030 &  &  &  &  &  &  \\
6228 & -0.195 & 0.063 & 1 & 0.047 &  &  &  &  &  &  \\
6560 & -0.127 & 0.116 & 2 & 0.105 &  &  &  &  &  &  \\
10705 & -0.311 & 0.271 & 2 & 0.024 &  &  &  &  &  &  \\
10741 & -0.265 & 0.031 & 1 & 0.011 &  &  &  &  &  &  \\
10753 & -0.065 & -0.005 & 1 & 0.028 &  &  &  &  &  &  \\
10968 & -0.262 & 0.253 & 2 & 0.242 &  &  &  &  &  &  \\
11180 & 0.007 & 0.090 & 2 & 0.044 &  &  &  &  &  &  \\
11203 & -0.265 & 0.204 & 2 & 0.239 &  &  &  &  &  &  \\
11273 & -0.168 & 0.125 & 2 & 0.029 &  &  &  &  &  &  \\
11281 & -0.130 & 0.104 & 2 & 0.866 &  &  &  &  &  &  \\
11294 & -0.196 & 0.047 & 1 & 0.070 &  &  &  &  &  &  \\
11305 & -0.085 & 0.026 & 1 & 0.050 &  &  &  &  &  &  \\
11306 & -0.082 & 0.219 & 2 & 0.023 &  &  &  &  &  &  \\
11317 & -0.252 & 0.038 & 1 & 1.000 & 0.83 & 0.42 & 1.5 & 0.249 & 603 & 6.64 \\
11425 & -0.147 & 0.202 & 2 & 0.028 &  &  &  &  &  &  \\
11455 & -0.351 & 0.100 & 1 & 0.910 &  &  &  &  &  &  \\
11585 & -0.201 & 0.056 & 1 & 0.329 &  &  &  &  &  &  \\
11750 & -0.231 & 0.185 & 2 & 0.060 &  &  &  &  &  &  \\
11806 & -0.088 & 0.181 & 2 & 0.137 &  &  &  &  &  &  \\
11821 & -0.316 & 0.024 & 1 & 0.010 &  &  &  &  &  &  \\
11888 & -0.245 & 0.283 & 2 & 0.221 &  &  &  &  &  &  \\
11918 & -0.166 & 0.024 & 1 & 0.030 &  &  &  &  &  &  \\
11942 & -0.141 & 0.106 & 2 & 0.053 &  &  &  &  &  &  \\
12115 & -0.339 & 0.280 & 2 & 0.015 &  &  &  &  &  &  \\
12253 & -0.268 & 0.194 & 2 & 0.138 &  &  &  &  &  &  \\
12309 & -0.217 & 0.193 & 2 & 0.257 &  &  &  &  &  &  \\
12319 & -0.267 & 0.058 & 1 & 0.093 &  &  &  &  &  &  \\
12322 & -0.228 & 0.205 & 2 & 0.040 &  &  &  &  &  &  \\
12363 & -0.112 & 0.039 & 1 & 0.024 &  &  &  &  &  &  \\
12468 & -0.260 & 0.011 & 1 & 0.024 &  &  &  &  &  &  \\
12517 & -0.186 & 0.261 & 2 & 0.046 &  &  &  &  &  &  \\
12646 & -0.415 & 0.165 & 1 & 0.047 &  &  &  &  &  &  \\
12658 & -0.198 & 0.071 & 1 & 1.000 & 0.83 & 0.40 & 0.659 & 0.073 & 176 & 14.7 \\
12833 & -0.263 & 0.274 & 2 & 0.030 &  &  &  &  &  &  \\
13019 & -0.191 & 0.089 & 1 & 0.053 &  &  &  &  &  &  \\
13112 & -0.299 & 0.221 & 2 & 0.802 &  &  &  &  &  &  \\
13174 & -0.153 & 0.016 & 1 & 0.046 &  &  &  &  &  &  \\
13438 & 0.000 & 0.110 & 2 & 1.000 & 0.82 & 0.35 & 0.0676 & 0.022 & 5.93 & 123 \\
13521 & -0.195 & 0.059 & 1 & 0.013 &  &  &  &  &  &  \\
13556 & -0.310 & 0.080 & 1 & 0.045 &  &  &  &  &  &  \\
13739 & -0.417 & 0.279 & 2 & 0.090 &  &  &  &  &  &  \\
13768 & -0.073 & 0.029 & 1 & 0.012 &  &  &  &  &  &  \\
13808 & -0.334 & 0.319 & 2 & 0.695 & 0.83 & 0.21 & 4.13 & 0.112 & 3e+03 & 1.24 \\
13816 & -0.541 & 0.165 & 1 & 0.834 & 0.83 & 0.11 & 0.864 & 0.166 & 302 & 3.19 \\
13824 & -0.247 & 0.243 & 2 & 0.168 &  &  &  &  &  &  \\
14175 & -0.209 & 0.060 & 1 & 0.057 &  &  &  &  &  &  \\
14302 & -0.091 & 0.006 & 1 & 0.040 &  &  &  &  &  &  \\
14465 & -0.259 & 0.268 & 2 & 0.052 &  &  &  &  &  &  \\
14601 & -0.331 & 0.071 & 1 & 0.047 &  &  &  &  &  &  \\
14789 & -0.059 & -0.003 & 1 & 0.075 &  &  &  &  &  &  \\
14815 & -0.248 & 0.250 & 2 & 0.030 &  &  &  &  &  &  \\
14830 & -0.404 & 0.358 & 2 & 0.022 &  &  &  &  &  &  \\
15012 & -0.035 & -0.002 & 1 & 0.074 &  &  &  &  &  &  \\
15013 & -0.280 & 0.272 & 2 & 0.104 &  &  &  &  &  &  \\
15069 & -0.233 & 0.274 & 2 & 0.016 &  &  &  &  &  &  \\
15101 & -0.212 & 0.203 & 2 & 0.047 &  &  &  &  &  &  \\
15165 & -0.157 & 0.202 & 2 & 0.007 &  &  &  &  &  &  \\
15182 & -0.144 & 0.014 & 1 & 0.018 &  &  &  &  &  &  \\
15293 & -0.536 & 0.158 & 1 & 1.000 & 0.82 & 0.49 & 1.64 & 0.477 & 669 & 7.16 \\
15382 & -0.146 & 0.023 & 1 & 0.579 &  &  &  &  &  &  \\
15422 & -0.143 & 0.234 & 2 & 0.041 &  &  &  &  &  &  \\
15482 & -0.284 & 0.077 & 1 & 1.000 &  &  &  &  &  &  \\
15528 & -0.300 & 0.267 & 2 & 0.052 &  &  &  &  &  &  \\
20774 & -0.270 & 0.176 & 2 & 0.042 &  &  &  &  &  &  \\
21050 & -0.202 & 0.211 & 2 & 0.167 &  &  &  &  &  &  \\
21060 & -0.084 & -0.009 & 1 & 0.129 &  &  &  &  &  &  \\
21131 & -0.039 & 0.036 & 1 & 0.115 &  &  &  &  &  &  \\
21189 & -0.158 & 0.178 & 2 & 0.005 &  &  &  &  &  &  \\
21232 & -0.142 & -0.006 & 1 & 1.000 &  &  &  &  &  &  \\
21271 & -0.027 & -0.012 & 1 & 0.051 &  &  &  &  &  &  \\
21272 & -0.252 & 0.296 & 2 & 0.039 &  &  &  &  &  &  \\
21273 & -0.282 & 0.221 & 2 & 0.072 &  &  &  &  &  &  \\
21292 & -0.113 & 0.183 & 2 & 0.047 &  &  &  &  &  &  \\
21707 & -0.249 & 0.240 & 2 & 0.011 &  &  &  &  &  &  \\
21918 & -0.275 & 0.264 & 2 & 0.008 &  &  &  &  &  &  \\
21921 & -0.101 & 0.021 & 1 & 0.999 &  &  &  &  &  &  \\
22325 & -0.099 & 0.166 & 2 & 0.042 &  &  &  &  &  &  \\
22396 & -0.011 & -0.008 & 1 & 0.022 &  &  &  &  &  &  \\
22401 & -0.130 & 0.093 & 2 & 0.163 &  &  &  &  &  &  \\
22488 & -0.317 & 0.262 & 2 & 0.320 &  &  &  &  &  &  \\
22686 & -0.376 & 0.252 & 2 & 0.029 &  &  &  &  &  &  \\
22751 & -0.556 & 0.120 & 1 & 1.000 & 0.83 & 0.30 & 0.631 & 0.027 & 173 & 11.4 \\
23045 & -0.255 & 0.218 & 2 & 0.051 &  &  &  &  &  &  \\
23175 & -0.200 & 0.022 & 1 & 1.000 & 0.83 & 0.12 & 0.491 & 0.124 & 129 & 5.78 \\
23330 & -0.030 & 0.001 & 1 & 0.010 &  &  &  &  &  &  \\
23342 & -0.101 & -0.005 & 1 & 0.465 &  &  &  &  &  &  \\
23375 & -0.232 & 0.246 & 2 & 0.487 &  &  &  &  &  &  \\
23452 & -0.257 & 0.241 & 2 & 1.000 & 0.82 & 0.20 & 0.69 & 0.056 & 206 & 7.01 \\
23461 & -0.181 & 0.265 & 2 & 0.040 &  &  &  &  &  &  \\
23519 & -0.212 & 0.034 & 1 & 0.046 &  &  &  &  &  &  \\
23640 & -0.198 & 0.155 & 2 & 0.044 &  &  &  &  &  &  \\
24190 & -0.140 & 0.271 & 2 & 0.007 &  &  &  &  &  &  \\
24416 & -0.276 & 0.253 & 2 & 0.012 &  &  &  &  &  &  \\
24524 & -0.098 & -0.003 & 1 & 0.220 &  &  &  &  &  &  \\
24592 & -0.151 & 0.033 & 1 & 0.045 &  &  &  &  &  &  \\
24594 & -0.355 & 0.352 & 2 & 0.024 &  &  &  &  &  &  \\
24684 & -0.272 & 0.400 & 2 & 0.028 &  &  &  &  &  &  \\
24753 & -0.007 & 0.005 & 1 & 0.030 &  &  &  &  &  &  \\
24803 & -0.149 & 0.037 & 1 & 0.031 &  &  &  &  &  &  \\
24832 & -0.266 & 0.258 & 2 & 0.040 &  &  &  &  &  &  \\
24875 & -0.231 & 0.269 & 2 & 0.030 &  &  &  &  &  &  \\
25058 & -0.185 & 0.041 & 1 & 0.037 &  &  &  &  &  &  \\
25322 & -0.317 & 0.076 & 1 & 0.799 &  &  &  &  &  &  \\
\end{longtable}
\tablefoot{\tablefoottext{a}{Assuming the orbit is oriented edge-on (i.e.~$i=90\,{\rm deg}$).}}
}
\end{table*}

\begin{acknowledgements}
SK and NB gratefully acknowledge funding from a European Research Council consolidator grant (ERC-CoG-646928- Multi-Pop).
BG, SD, TOH, and ML  acknowledge  funding  from  the Deutsche Forschungsgemeinschaft (grant DR 281/35-1 and KA  4537/2-1)  and  from  the  German  Ministry  for  Education and Science (BMBF Verbundforschung) through grants 05A14MGA, 05A17MGA, 05A14BAC, and 05A17BAA.
NB gratefully acknowledges financial support from the Royal Society in the form of a University Research Fellowship.
JB acknowledges support by FCT/MCTES through national funds (PIDDAC) by grant UID/FIS/04434/2019 and through Investigador FCT Contract No. IF/01654/2014/CP1215/CT0003.
\end{acknowledgements}

%
\bibliographystyle{aa} 
\bibliography{ngc3201_multipop_binaries.bib} 
%



\end{document}